\documentclass[final,5p,twocolumn]{elsarticle}

\usepackage{graphicx}          
\usepackage[dvips]{epsfig}    
\usepackage{xcolor}

\usepackage{amssymb}
\usepackage{amsthm}

\usepackage{amsmath}          

\usepackage{algorithmic}

\journal{for ASME publication (accepted)}

\begin{document}

\begin{frontmatter}

\title{Extended fractional-order Jeffreys model of viscoelastic hydraulic cylinder} 

\author{Michael Ruderman}
\ead{michael.ruderman@uia.no}


\address{University of Agder, 4604-Norway}

\begin{abstract}
A novel modeling approach for viscoelastic hydraulic cylinders,
with negligible inertial forces, is proposed, based on the
extended fractional-order Jeffreys model. Analysis and physical
reasoning for the parameter constraints and order of the
fractional derivatives are provided. Comparison between the
measured and computed frequency response functions and time domain
transient response argues in favor of the proposed four-parameter
fractional-order model.
\end{abstract}

\end{frontmatter}

\section{Introduction}
\label{sec:1}

In viscoelasticity, fractional-order models have since long been
recognized as being more accurate and flexible than other
approaches, in addition to having a lower dimension of parameter
space than their integer-order counterparts. Analysis of the
various properties of viscoelastic behavior, such as creep,
relaxation, viscosity and initial conditions, can be found, e.g.,
in \cite{heymans2006,mainardi2011}. It is also worth recalling
that one of the widespread formulations of the fractional-order
differential operator, namely the Caputo derivative, has been
elaborated in the context of a viscoelastic stress-strain relation
and memory mechanism associated with the initial conditions (see,
e.g., \cite{caputo1971} for details).

It is not only viscoelastic solids that have benefited from
fractional-order modeling. Soft biological tissues have also been
studied experimentally and identified by means of the fractional
calculus in viscoelasticity \cite{meral2010}; a more recent review
of fractional calculus in modeling biological phenomena can be
found in \cite{ionescu2017}. Other examples of the complex impact
and retardation dynamics, addressed with the help of
fractional-order modeling, can be found when analyzing, e.g.,
backlash \cite{machado2013} or piezoelectric creep \cite{liu2013}.
Fractional-order modeling has been shown to be efficient and has
also been identified in experiments describing the transient
behavior of supercapacitors \cite{freeborn2013}. Besides the
above-mentioned examples, significant efforts in fractional-order
calculus of dynamic systems should be credited to fractional-order
controllers, though not directly related to the recent study. For
exemplary overviews of the tuning of fractional-order controllers
for industry applications and fractional
proportional-integral-derivative (PID) controllers, we refer to
e.g. \cite{monje2008} and \cite{shah2016}, respectively.

The coupled hydrodynamical and mechanical response of hydraulic
cylinders, especially with lower acceleration rates, can be
challenging when it comes to accurate description by Newton's
classical lumped-mass laws of motion. Instead, the viscoelastic
behavior of the dashpot type can be seen to be beneficial. The
integer-order Jeffreys \cite{joseph1990} viscoelastic model takes
into account not only the dashpot, which is Newtonian fluid, but
also the Kelvin-Voigt viscoelastic structure, which appears to be
suitable for mimicking the compressibility of a hydraulic medium.
While the classical Jeffreys model is well established and
suitable for non-Newtonian fluids, generally allowing for
relaxation of stress and strain at different rates, its
integer-order dynamics can be suboptimal for different-type
hydro-mechanical couplings. Note that an integer order in the
Jeffreys model implies a chain of two integrators, which would
restrict the phase response of transfer characteristics by --180
deg at higher frequencies. Involvement of fractional-order
dynamics, in contrast, provides additional flexibility in shaping
the frequency characteristics of a hydro-mechanical transducer. At
the same time, it allows the dimension of the free model
parameters to be kept as low as possible. This is quite the
opposite of lumped-parameter rheological modeling approaches,
where more sophisticated structures of parallel and serial
connections of the viscoelastic elements are involved, in order to
approximate the input-output transfer characteristics of interest.

In this brief, we make use of fractional-order viscoelastic
behavior to describe the principal dynamics of standard
dashpot-type hydraulic cylinders, with a low-parameter and
low-order model that can still accurately predict the frequency
response characteristics. The proposed solution follows the
Jeffreys \cite{joseph1990} viscoelastic model (though extending
this by an additional stiffness), generalizing to fractional
derivatives, and setting physically reasonable constraints on the
parameters and differential order. While providing all necessary
preliminaries of the fractional-order calculus (see Section 2),
for the basics on applied hydraulics we refer to
\cite{jelali2012}. The main contribution of this paper is in
Section 3, and the experimental evaluation and discussion are
provided in Section 4.

\section{Fractional differentiation}
\label{sec:2}

In accordance with the seminal literature on fractional-order
calculus, e.g., \cite{oldham1974,samko1993,podlubny1998}, we will
use a fractional $\alpha$-order operator on the $a$ and $t$
limits, defined by
\begin{equation}\label{eq:2:1}
    _{a}D_{t}^{\alpha} = \left\{%
\begin{array}{ll}
    \frac{d^{\alpha}}{ d t^{\alpha}} & \hbox{for } \alpha > 0,
    \\[0.2cm]
    1 & \hbox{for } \alpha =0, \\[0cm]
    \int \limits^{t}_{a} dt^{-\alpha} & \hbox{for } \alpha <0. \\
\end{array}%
\right.
\end{equation}
For the sake of practical relevance, we confine ourselves to real
fractional orders, i.e., $\alpha \in \mathbb{R} $, often denoted
as a \emph{non-integer} differentiation, correspondingly
non-integer integration. For the same reasoning as in practical
(engineering) applications, we will consider zero initial time,
i.e., $a=0$.

The classical definition of the Riemann-Liouville
$\alpha$-derivative of a continuous function $f(t)$, with $n-1
\leq \alpha < n$, where $n$ is an integer, is given by
\begin{equation}\label{eq:2:2}
    _{0}D_{t}^{\alpha}f(t)=\frac{1}{\Gamma(n-\alpha)}
    \frac{d^{n}}{dt^{n}}
    \int \limits_{0}^{t}\frac{f(\tau)}{(t-\tau)^{\alpha-n+1}} \,
    d\tau.
\end{equation}
Here, $\Gamma (\cdot)$ is the gamma function. We should point out
that solving differential equations in terms of Riemann-Liouville
derivatives \eqref{eq:2:2} requires the initial conditions
\begin{equation}\label{eq:2:3}
    \bigl[_{0}D_{t}^{\alpha-k}f(t)\bigr]_{t\rightarrow 0} = c_{k} \; \hbox{ for
    } \; k=1,2,\ldots,n
\end{equation}
to be known and determined accordingly. That is, one needs the
initial values of the $(\alpha-k)$-fractional derivatives of the
function $f(t)$. This is particularly evident when considering the
Laplace transform $F(s)=\mathcal{L}\bigl\{ f(t) \bigr\}$ of the
Riemann-Liouville fractional derivative (cf. \cite{podlubny1998}),
given by
\begin{equation}\label{eq:2:4}
    \mathcal{L} \bigl\{ _{0}D_{t}^{\alpha}f(t) \bigr\} = s^{\alpha}F(s)-\sum
    _{k=0}^{n-1}s^{k} \bigl[ _{0}D_{t}^{\alpha-k-1}f(t) \bigr]_{t=0} \,.
\end{equation}

The Gr\"unwald-Letnikov definition (cf. \cite{podlubny1998})
\begin{equation}\label{eq:2:5}
    _{0}D_{t}^{\alpha}f(t) = \underset {h\rightarrow 0} \lim \, h^{-\alpha}\sum
    _{i=0}^{[th^{-1}]}(-1)^{i} \left(%
\begin{array}{c}
  \alpha \\
  i \\
\end{array}%
\right) f(t-ih)
\end{equation}
of the fractional-order derivative is valid for any $\alpha \in
\mathbb{R}$ and is particularly suitable for numerical
implementations, since it constitutes the limit of the difference
quotient
\begin{equation}\label{eq:2:6}
    \Delta_{
    h}^{\alpha}f(t) \approx \, _{0}D_{t}^{\alpha}f(t),
\end{equation}
with the time step $h\rightarrow 0$. In \eqref{eq:2:5}, the
operator $[x]$ means the integer part of $x$, while $i$ is the
index of the discrete time series of $t$. The binomial
coefficients, which are sign-alternating and summarized as
\begin{equation}\label{eq:2:7}
    w_{i}^{(\alpha)}=(-1)^{i}  \left(%
\begin{array}{c}
  \alpha \\
  i \\
\end{array}%
\right) \; \hbox{ for } \;  i = 0,1,2,\ldots,
\end{equation}
can be evaluated recursively (cf. \cite{podlubny1998}) by
\begin{equation}\label{eq:2:7}
    w_{0}^{(\alpha)}=1; \; w_{i}^{(\alpha)} = \Bigl( 1-\frac{\alpha
    +1}{i} \Bigr) w_{i-1}^{(\alpha)}  \; \hbox{ for } \; i = 1,2,3,\ldots \; .
\end{equation}
Further, we will also make use of the fact that for zero initial
conditions, the Laplace transform is given by
\begin{equation}\label{eq:2:8}
    \mathcal{L}_{0} \bigl\{ _{0}D_{t}^{\alpha}f(t) \bigr\} = s^{\alpha}F(s),
\end{equation}
and we will write the Fourier transform as
\begin{equation}\label{eq:2:9}
    \mathcal{F} \bigl\{ _{0}D_{t}^{\alpha} f(t) \bigr\} =
    (j\omega)^{\alpha}F(\omega).
\end{equation}
Note that for $\mathcal{F}$, the magnitude and phase responses are
conventionally given as in the case of $\alpha \in \mathbb{Z}$.

\section{Modeling}
\label{sec:3}

Our starting point is the standard dashpot element used in
viscoelasticity for analyzing and representing an ideal linear
viscous fluid. A constantly applied stress $\sigma$ produces a
constant strain rate so that
\begin{equation}\label{eq:3:1}
    \sigma=\mu\frac{d}{dt}\varepsilon,
\end{equation}
where $\mu$ is the coefficient of viscosity of a Newtonian fluid.
Note that a linear dashpot, as a basic mechanical element, is
approximating a simple piston cylinder, with one degree of freedom
$x$, where the stress equivalent of the hydraulic pressure is
replaced by the applied force  $\tau$. Correspondingly, the strain
rate is equivalent to the rate of displacement so that, in the
Laplace domain, one obtains
\begin{equation}\label{eq:3:2}
    x=\mu^{-1}s^{-1}\tau.
\end{equation}
Obviously, \eqref{eq:3:2} constitutes a free integrator dynamic,
this way yielding a simple first-order lumped model of the viscous
driven motion when inertial effects are neglected. Note that the
latter is justified for multiple hydraulic cylinders, deployed as
the actuators, in which the viscoelastic forces largely dominate
over the inertial.

Next, we elaborate whether the hydraulic cylinder force, induced
by the pressure difference, itself undergoes the dynamic
transients to be captured by viscoelastic behavior. The
stress-strain relation of the standard linear solid model, also
denoted as Zener model, is give by
\begin{equation}\label{eq:3:3}
    E \Bigl[ \frac{1}{\phi}\frac{d}{dt} + 1 \Bigr] \varepsilon(t) = \Bigl[ \frac{1}{\varphi}\frac{d}{dt} + 1 \Bigr]\sigma(t),
\end{equation}
where $E$ is a suitable elastic modulus. The positive constants
$\phi^{-1} >\varphi^{-1}$ refer to the retardation and relaxation
times (see, e.g., \cite{joseph1990}). It can be noted that $\phi=
\varphi$ simplifies \eqref{eq:3:3} to a purely elastic material,
i.e., $\sigma=E \varepsilon$, while \eqref{eq:3:3} includes
equally the Maxwell and Kelvin-Voigt models of the viscoelastic
fluid, correspondingly solid, as limiting cases \cite{caputo1971}.
The otherwise unequal and non-zero time constants of retardation
and relaxation shape the frequency characteristics of a standard
linear solid within $\phi < \omega< \varphi$, while keeping it at
different constant levels for $\omega\rightarrow 0$ and $\omega
\rightarrow \infty$ limits. Since the dynamic behavior
\eqref{eq:3:3} has zero relative degree, its input-output transfer
characteristics can also be considered as entirely unitless. For
approaching a viscoelastic fluid in the closed hydraulic circuit
(of a powered cylinder) as a quasi-solid medium, the model
\eqref{eq:3:3} can be used to describe its dynamic
compressibility. Thus, the applied stress $\sigma$ appears in
equivalence to the supplied pressure difference, while the strain
$\varepsilon$ mimics a stiff kinematic excitation, and therefore
force, acting on the piston interface. In this equivalence, the
elastic modulus appears as the bulk modulus \cite{jelali2012} of
the hydraulic medium. The retardation time approaches the time
constant of the effective force response to a stepwise change of
the differential pressure. Similarly, the relaxation time relates
to the time constant of the pressure response to an instantaneous
change in the piston stroke and thus force on the piston
interface.

\subsection{Jeffreys model} \label{sec:3:sub:1}

\begin{figure}[!h]
\centering
\includegraphics[width=0.55\columnwidth]{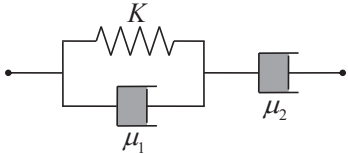}
\caption{Jeffreys viscoelastic model.} \label{fig:1}
\end{figure}

The Jeffreys model \cite{joseph1990} comprises, essentially, the
Kelvin-Voigt and dashpot elements connected in series (see
structural arrangement in Figure \ref{fig:1}).
The overall strain (or relative displacement) rate is a
superposition of both
\begin{equation}\label{eq:3:4}
    \frac{d}{dt}\varepsilon=\frac{d}{dt}\varepsilon
    _{1}+\frac{d}{dt}\varepsilon _{2},
\end{equation}
while the stress (corresponding force) in both elements is the
same, meaning
\begin{equation}\label{eq:3:5}
    \sigma=K\varepsilon _{1}+\mu _{1}\frac{d}{dt}\varepsilon
    _{1}=\mu _{2}\frac{d}{dt}\varepsilon_{2}.
\end{equation}
Combining \eqref{eq:3:4} and \eqref{eq:3:5}, and eliminating
$\varepsilon_{1}$ and $\varepsilon _{2}$, results in the Jeffreys
model of the form (see \cite{joseph1990} for details):
\begin{equation}\label{eq:3:6}
    \mu_{2} \Bigl[ \frac{\mu _{1}}{K}\frac{d^{2}}{dt^{2}}+1 \Bigr] \varepsilon(t) =
    \Bigl[ \frac{\mu_{1}+\mu_{2}}{K}\frac{d}{dt}+1 \Bigr] \sigma(t).
\end{equation}
Here, the relaxation and retardation times are given by
$(\mu_{1}+\mu_{2})K^{-1}$ and $\mu_{1}K^{-1}$, correspondingly.
Note that, as in the case of the Zener model, elimination (set to
zero) of $\mu_{1}$ or $\mu_{2}$ simplifies the Jeffreys model
\eqref{eq:3:6} to the Maxwell or Kelvin-Voigt model, respectively.

\subsection{Fractional-order formulation} \label{sec:3:sub:2}

Allowing for the fractional-order derivatives and considering the
input-output pair of differential pressure (equivalent to
hydraulic force) and relative displacement, correspondingly, the
fractional-order Jeffreys model, in the Laplace domain, is written
as
\begin{equation}\label{eq:3:7}
    \mu s^{\gamma} \bigl(\lambda_{ 2}s^{\alpha}+1 \bigr) x(s) = \bigl( \lambda
    _{1}s^{\beta}+1 \bigr) \tau(s).
\end{equation}
Note that a similar modified Jeffreys model has also been
proposed, though for viscoelastic fluids only, in \cite{song1998}.
It can be recognized that for the same time constants $\lambda
_{1}=\lambda _{2}$ and fractional orders $\alpha=\beta$, the
proposed model \eqref{eq:3:7} is reduced to the simple viscous
dashpot, cf. with \eqref{eq:3:2}.

Unlike the original Jeffreys model (cf. \cite{joseph1990}) and the
modified one \cite{song1998}, the inequality $\lambda_{ 2}>\lambda
_{1}$ of parameters is required for \eqref{eq:3:7}. Note that this
implies the lag characteristics of the $x(s)/\tau (s)$ transfer
function, which means a phase-lowering region and amplitude drop
within $\lambda_{2}^{-1}< \omega < \lambda_{1}^{-1}$. From a
structural viewpoint, it would require an additional spring that
would lead to a serial connection of the Zener model with a
viscous dashpot (see above in Section \ref{sec:3}), while the
dynamics order and relative degree remain unchanged this way. The
viscoelastic damping properties of the Zener model \eqref{eq:3:3}
argue in favor of the lag characteristics of \eqref{eq:3:7}.
Indeed, the force propagation through the hydraulic medium, from
the differential pressure source to the effective piston force, is
weakened for higher frequencies and additionally lagged for a
certain frequency range. Here, we recall that for the Zener model
to be dissipative and, therefore, physically reasonable for real
materials (and corresponding media), the thermodynamic constraints
of the $\alpha=\beta$ parameters should be additionally satisfied.
This has been observed previously in multiple rheological studies
and also theoretically proved in \cite{bagley1986}.

While the general form \eqref{eq:3:7} allows for all derivatives
to be fractional-order, we further require $\gamma=1$ to avoid
violating causality and physical reasoning of the
force-displacement transfer characteristics. This is analyzed
below in the context of the impulse response and its steady-state
(final) value. The above physical constraints on the time
constants and differential orders lead us to the overall
four-parameter linear fractional-order model \eqref{eq:3:7} of
viscoelastic hydraulic cylinders with negligible inertial terms.

\subsection{Initial value and impulse response}
\label{sec:3:sub:3}

Next, we need to clarify the initial conditions for fractional
differential equation \eqref{eq:3:7}. The initial condition for
the integer-order dashpot, which is the relative displacement at
$t=0$, can be assumed to be zero without loss of generality. The
remaining fractional-order dynamics of the Zener model require the
single initial condition $_{0}D_{t}^{\alpha-1}x(t)$ (cf.
\cite{heymans2006}), provided the input differential pressure
(corresponding actuation force) is a known exogenous value. It has
been shown in \cite{heymans2006} that for a physically reasonable
(i.e., continuous) loading, or even in the case of a step
discontinuity, zero initial conditions for the strain dynamics
apply. Non-zero initial conditions will only be valid in the case
of a $B\, \delta(t)$ stress impulse (cf. \cite{heymans2006}),
resulting in
\begin{equation}\label{eq:3:8}
    \Bigl[ \lambda_{2} \, _{0} D_{t}^{\alpha -1}x(t) \Bigr]_{t \rightarrow 0} =
    B.
\end{equation}

\begin{figure}[!h]
\centering
\includegraphics[width=0.95\columnwidth]{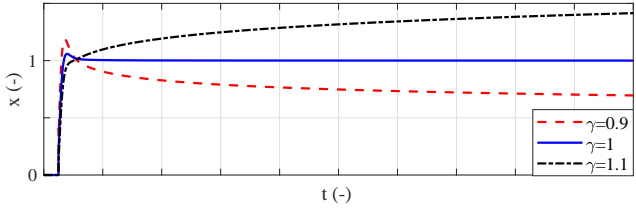}
\caption{Impulse response for various order $\gamma =\{0.9,\, 1,\,
1.1\}$.} \label{fig:2}
\end{figure}
We now have to take a closer look at the impulse response of
\eqref{eq:3:7}, in terms of the final value problem from which the
$\gamma=1$ constraint is enlightened. Rewriting \eqref{eq:3:7} in
the time domain and integrating both sides with respect to
$dt^{\gamma}$ results in
\begin{equation}\label{eq:3:9}
    \mu \lambda_{2} \, _{0}D_{t}^{\alpha} x(t)+\mu
    x(t) =\, _{0}D_{t}^{-\gamma} \bigl( \lambda_{1} \, _{0}D_{t}^{\beta} \tau
    (t)+\tau (t) \bigr).
\end{equation}
The steady state of \eqref{eq:3:9} can be obtained via the final
value theorem, since for $\alpha,\beta > 0$, both differential
terms in \eqref{eq:3:9} vanish for $t\rightarrow \infty$. Solving
the final value problem for $x(t)$ this way yields
\begin{equation}\label{eq:3:10}
    \bigl[ x(t) \bigr]_{t\rightarrow \infty} = \mu ^{-1} \int\limits ^{t} _{0} \tau
    (t)dt^{\gamma}.
\end{equation}
For the applied Dirac impulse $\delta(t)$, the integral solution
\eqref{eq:3:10} can be directly evaluated (see, e.g.,
\cite{podlubny1998}), given as
\begin{equation}\label{eq:3:11}
    _{0}D_{t}^{\eta} \delta(t) = \frac{t^{-\eta-1}}{\Gamma(-\eta)}.
\end{equation}
It can be shown that for $\eta < 0$, the value of \eqref{eq:3:11}
remains, $\forall \: t > 0$, constant and equal to one for
$\eta=-1$ only. Otherwise, it converges to zero for $-1< \eta <0$
and diverges for $\eta < -1$. For the final value \eqref{eq:3:10},
it means that a constant non-zero impulse response of the relative
displacement can only be achieved when $\gamma =1$. Indeed, the
free integrator in \eqref{eq:3:7} requires a constant finite
displacement as a result of a force impulse which has finite
energy content. An illustrative numerical example of the
\eqref{eq:3:7} model response to the input impulse is shown in
Figure \ref{fig:2} for $\gamma =\{0.9,\, 1,\, 1.1\}$.

\section{Experimental evaluation and discussion}
\label{sec:4}

\begin{figure}[!h]
\centering
\includegraphics[width=0.95\columnwidth]{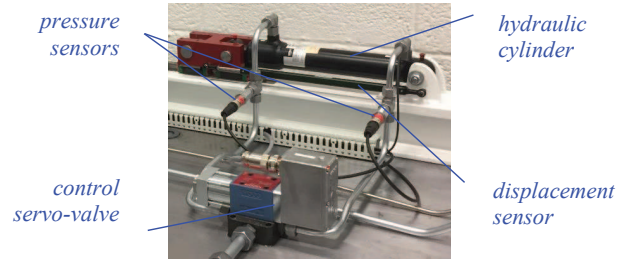}
\caption{Experimental setup (laboratory view) of the
valve-controlled hydraulic cylinder with displacement and pressure
sensors.} \label{fig:2new}
\end{figure}

Experimental evaluation of the above model was made with the data
recorded from a standard linear-stroke hydraulic cylinder in a
laboratory setting (see Figure \ref{fig:2new}). The standard
one-side-rod hydraulic cylinder, with a full cross-section of 25
mm, is actuated via a 4/3 servovalve with the operational supply
pressure set to 100 bar. The relative displacement $x(t)$ is
directly measured by an absolute resistor-based linear
potentiometer, while the total effective stroke is about 200 mm.
Two pressure sensors, with a measurement range up to 250 bar, are
installed on the fittings, close to both cylinder chambers. The
differential pressure $\tau (t)$ is directly obtained from the
measurements. All experimental data are generated in an open-loop
control manner by designing and feeding the reference control
signal for the servovalve for both frequency domain and time
domain measurements. The frequency domain data, used below for
identification, were collected as single steady-state points for
the set of harmonic excitations, and that using averaged sine
waves of different frequencies between 0.005 Hz and 1.6 Hz.
Further details of the experimental setup can be found in
\cite{pasolli2018}.

\subsection{Frequency response function} \label{sec:4:sub:1}

\begin{figure}[!h]
\centering
\includegraphics[width=0.98\columnwidth]{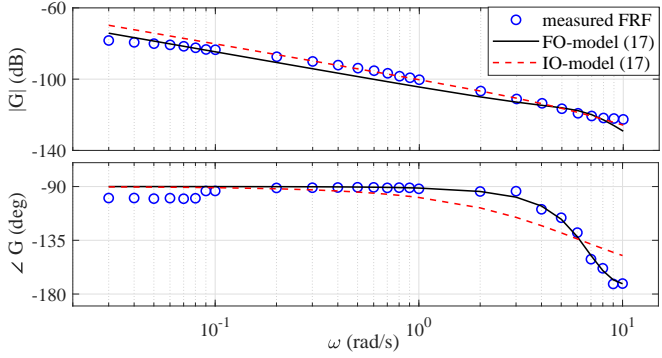}
\caption{Measured and \eqref{eq:3:7}-modeled frequency response
function; least-squares best fit for fraction- and integer-order
derivatives.} \label{fig:3}
\end{figure}

The measured frequency response function
$G(j\omega)=x(j\omega)/\tau(j\omega)$ is used for the
least-squares fit of the fractional-order (FO) model
\eqref{eq:3:7}. Note that both the amplitude response in dB and
phase response in deg have been incorporated into the objective
function. Given the comparable range of dB and deg units, the
amplitude and phase response errors are equally included in the
overall objective function of numerical minimization. For the sake
of comparison, the least-squares best fit has also been found
(from the same experimental data) for the integer-order (IO) model
with the structure as in \eqref{eq:3:7}. The measured and
both-ways modeled frequency response functions are shown opposite
to each other in Figure \ref{fig:3}. The fractional-order model
coincides with the measurements, while the integer-order one
misses the phase response.
The identified parameters of the proposed model \eqref{eq:3:7} are
$\lambda_1 = 0.013$, $\lambda_2 = 0.047$, $\alpha=\beta=1.571$,
$\mu = 171e3$.

\subsection{Time series} \label{sec:4:sub:2}

\begin{figure}[!h]
\centering
\includegraphics[width=0.98\columnwidth]{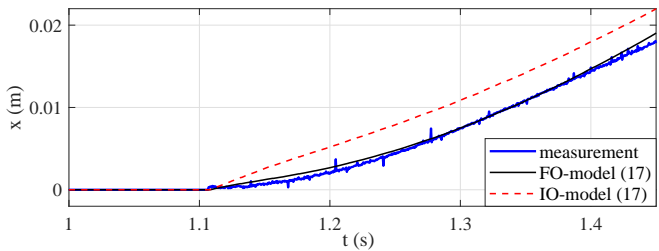}
\caption{Measured and \eqref{eq:3:7}-modeled time response to
slope input.} \label{fig:4}
\end{figure}

A similar observation can be made when comparing the time series
of the identified fractional- and integer-order models with the
initial phase of the measured response to the slope-shaped input
$\tau(t)$. From Figure \ref{fig:4}, one can see that only the
fractional-order model captures a lagged transient of the relative
displacement. Note that further progress of the measured response
(not shown in the figure) diverges from each model in all cases,
due to an inherent free integrator error. Yet the fractional-order
model reproduces the initial transition of the viscoelastic
response to a linearly increasing input force.

\subsection{Concluding remarks} \label{sec:4:sub:3}

The fraction-order Jeffreys model manifests as more accurate, the
opposite of its integer-order counterpart, when describing the
dynamic behavior of a viscoelastic hydraulic cylinder. This
becomes particularly visible in the frequency domain for the phase
response and, as a logical consequence, within the transient shape
in the time domain. The identified four-parameter model
\eqref{eq:3:7} takes into account the physical reasoning of
force-displacement transfer characteristics and thermodynamic
constraints.

\section*{Acknowledgments}

This research received funding from the IS-DAAD program under RCN
project number 294835. Laboratory measurements performed by
Daniela Kapp are also acknowledged.

\bibliographystyle{elsarticle-num}        

\bibliography{references}

\begin{thebibliography}{10}
\expandafter\ifx\csname url\endcsname\relax
  \def\url#1{\texttt{#1}}\fi
\expandafter\ifx\csname urlprefix\endcsname\relax\def\urlprefix{URL }\fi
\expandafter\ifx\csname href\endcsname\relax
  \def\href#1#2{#2} \def\path#1{#1}\fi

\bibitem{heymans2006}
N.~Heymans, I.~Podlubny, Physical interpretation of initial conditions for
  fractional differential equations with riemann-liouville fractional
  derivatives, Rheol. Acta 45 (2006) 765--771.

\bibitem{mainardi2011}
F.~Mainardi, G.~Spada, Creep, relaxation and viscosity properties for basic
  fractional models in rheology, The European Physical Journal Special Topics
  193~(1) (2011) 133--160.

\bibitem{caputo1971}
M.~Caputo, F.~Mainardi, A new dissipation model based on memory mechanism, Pure
  and App. Geop. 91 (1971) 134--147.

\bibitem{meral2010}
F.~Meral, T.~Royston, R.~Magin, Fractional calculus in viscoelasticity: an
  experimental study, Communications in Nonlinear Science and Numerical
  Simulation 15~(4) (2010) 939--945.

\bibitem{ionescu2017}
C.~Ionescu, A.~Lopes, D.~Copot, J.~T. Machado, J.~Bates, The role of fractional
  calculus in modeling biological phenomena: A review, Communications in
  Nonlinear Science and Numerical Simulation 51 (2017) 141--159.

\bibitem{machado2013}
J.~T. Machado, Fractional order modelling of dynamic backlash, Mechatronics
  23~(7) (2013) 741--745.

\bibitem{liu2013}
Y.~Liu, J.~Shan, N.~Qi, Creep modeling and identification for piezoelectric
  actuators based on fractional-order system, Mechatronics 23~(7) (2013)
  840--847.

\bibitem{freeborn2013}
T.~J. Freeborn, B.~Maundy, A.~S. Elwakil, Measurement of supercapacitor
  fractional-order model parameters from voltage-excited step response, IEEE
  Journal on Emerging and Selected Topics in Circuits and Systems 3~(3) (2013)
  367--376.

\bibitem{monje2008}
C.~A. Monje, B.~M. Vinagre, V.~Feliu, Y.~Chen, Tuning and auto-tuning of
  fractional order controllers for industry applications, Control engineering
  practice 16~(7) (2008) 798--812.

\bibitem{shah2016}
P.~Shah, S.~Agashe, Review of fractional {PID} controller, Mechatronics 38
  (2016) 29--41.

\bibitem{joseph1990}
D.~D. Joseph, Fluid Dynamics of Viscoelastic Liquids, Springer, 1990.

\bibitem{jelali2012}
M.~Jelali, A.~Kroll, Hydraulic servo-systems: modelling, identification and
  control, Springer, 2012.

\bibitem{oldham1974}
K.~Oldham, J.~Spanier, The fractional calculus theory and applications of
  differentiation and integration to arbitrary order, Elsevier, 1974.

\bibitem{samko1993}
S.~G. Samko, A.~A. Kilbas, O.~I. Marichev, et~al., Fractional integrals and
  derivatives, Gordon and Breach Science, 1993.

\bibitem{podlubny1998}
I.~Podlubny, Fractional differential equations, Elsevier, 1998.

\bibitem{song1998}
D.~Y. Song, T.~Q. Jiang, Study on the constitutive equation with fractional
  derivative for the viscoelastic fluids--modified {Jeffreys} model and its
  application, Rheol. Acta 37 (1998) 512--517.

\bibitem{bagley1986}
R.~L. Bagley, P.~J. Torvik, On the fractional calculus model of viscoelastic
  behavior, J. of Rheology 30 (1986) 133--155.

\bibitem{pasolli2018}
P.~Pasolli, M.~Ruderman, Linearized piecewise affine in control and states
  hydraulic system: Modeling and identification, in: IEEE 44th Annual
  Conference of the Industrial Electronics Society, 2018, pp. 4537--4544.

\end{thebibliography}

\end{document}